# Demonstration of a scalable all-solid-state refrigerator exploiting diffusion geometries and limiting interfacial conductances at temperatures below 1 Kelvin

***Northrop Grumman Microelectronics Design and Applications Team**

Northrop Grumman Corp.

## Abstract

Solid-state refrigerators using Normal-metal/Insulator/Superconductor (NIS) junctions have previously demonstrated excellent electron cooling but limited ability to cool phonons. The energy gap of the superconductor is used as an energy filter to allow higher than average energy electrons to preferentially tunnel from the normal-metal through the insulator into the superconductor where they travel as quasi-particles. Typically, the heat is moved and work is done to deposit hot quasi-particles into a normal-metal quasi-particle trap for rejection to the next refrigeration stage. Realizing that (1) the quasi-particles flow diffusively, driven by a concentration gradient in the electric field-free superconductor, and (2) that the undesirable backwards leaking of heat from the hot-side trap can be reduced by engineering the geometry and materials at the superconductor to trap interface, enhanced cooling can be achieved. Fabrication of the refrigerator was accomplished using a tungsten and titanium-tungsten alloy as the cold-side normal-metal, aluminum oxide as the insulator, aluminum as the superconductor, and gold as the trap, with the cold-side NIS portion being attached to the hot-side gold trap by bump bonding. The refrigerator consisted of 1121 junction pairs, each pair being an SINIS unit, all electrically connected in series. Using this we have measured the effective *phonon* temperature of a 3.9 mm x 3.9 mm x 0.65 mm silicon chip driven down to 70 mK from a bath temperature of 120 mK, and down to 174 mK from a 271 mK rejection temperature (a cooling of -97 mK). This is the first demonstration of the sub 1 K cooling of an entire silicon chip using NIS junctions. This configuration is mechanically robust and scalable. A 3D simulation model was developed and matches the cooling data as a function of drive current. Extending this model shows that additional, tenable, material and geometry revisions may provide large cooling performance improvements.

## I. INTRODUCTION

Future computers using qubits will require cooling to sub-Kelvin temperatures. Currently $^{3}He/^{4}He$ dilution refrigerators and adiabatic demagnetization refrigerators (ADR) are used to keep these circuits at such low temperatures. While ADR does not use $^{3}He$, it has several drawbacks preventing its use in large-scale, general-purpose, applications. Notably, ADR requires cyclic magnetization and demagnetization which prevents continuous refrigeration without the use of multiple parallel refrigerators. ADR also produces less cooling power than equivalently-sized dilution refrigerators and it poses significant magnetic field risks to its load devices. The $^{3}He$ isotope used in dilution refrigerators is a scarce commodity meaning these systems are challenging to adapt to large scale circuits. Previously, a third type of cooler using the tunneling of electrons across a Normal-metal / Insulator / Superconductor (NIS) junction has been demonstrated to cool electrons down into the milli-Kelvin range[1,2,3,4] using circuits fabricated by micro-lithography, which can be well scaled as integrated circuits. To function as a refrigerator for superconducting computers, it must cool not just the electrons but also the lattice phonons. This is necessary to maintain a sufficiently long state lifetime for computation. Despite these challenges, several groups have accomplished NIS cooling of phonons by utilizing membranes and nano-bridges or nano-wires for thermal isolation[5,6,7,8,9]. Notably, using this style of thermal isolation, Mykkänen et al.[10] demonstrated the cooling of the phonon temperature, at a 244 mK bath temperature, to 161 mK in a 0.4 mm thick by 1 mm diameter silicon island suspended on micro-bridges. Lowell et al.[11] cooled a 1.9 $cm^3$ copper stage from 290 to 256 mK by putting the NIS junctions on a membrane and connecting the copper stage to the cooler using a wire bond. This last work also illustrates that to use an electron refrigerator to cool the phonons it is necessary to have good coupling between the phonons and electrons.[3] As will be discussed below, this can be accomplished by increasing the normal-metal volume and by choice of a normal-metal with a relatively large electron-phonon coupling coefficient, or both.

## II. EXPERIMENT

To achieve the coldest phonon temperatures at the highest rejection (hot side) temperatures two concepts are implemented in the current work. First, the superconductor, in which the injected quasi-particles diffuse, is an electric field-free region and, as such, the heat carried by the quasi-particles is moving the heat from the cold side to the hot side driven by their concentration gradient (“uphill” against the phonon temperature gradient). Succinctly put, the quasi-particle heat flow is governed by Fick’s Law rather than Ohm’s Law. Second, to limit the amount of hot side backward heat leakage, the contact area between hot side and cold side should be minimized.

These considerations brought about our choice to fabricate the cold and hot sides separately and join them together by Au-Au compression bump bonding[12] (Figure 1). The heat carrying quasi-particles are injected across the NIS junction, tunneling from the normal-metal into the superconductor. Our best results were obtained with junction areas 50 times larger than the bump contact. This requires that the injected quasi-particles must diffuse, without losing too much heat by recombination, to the Au trap[13] on the hot side. On reaching the trap, the quasi-particle actualize their I-V work as heat in addition to the heat removed from the cold side.

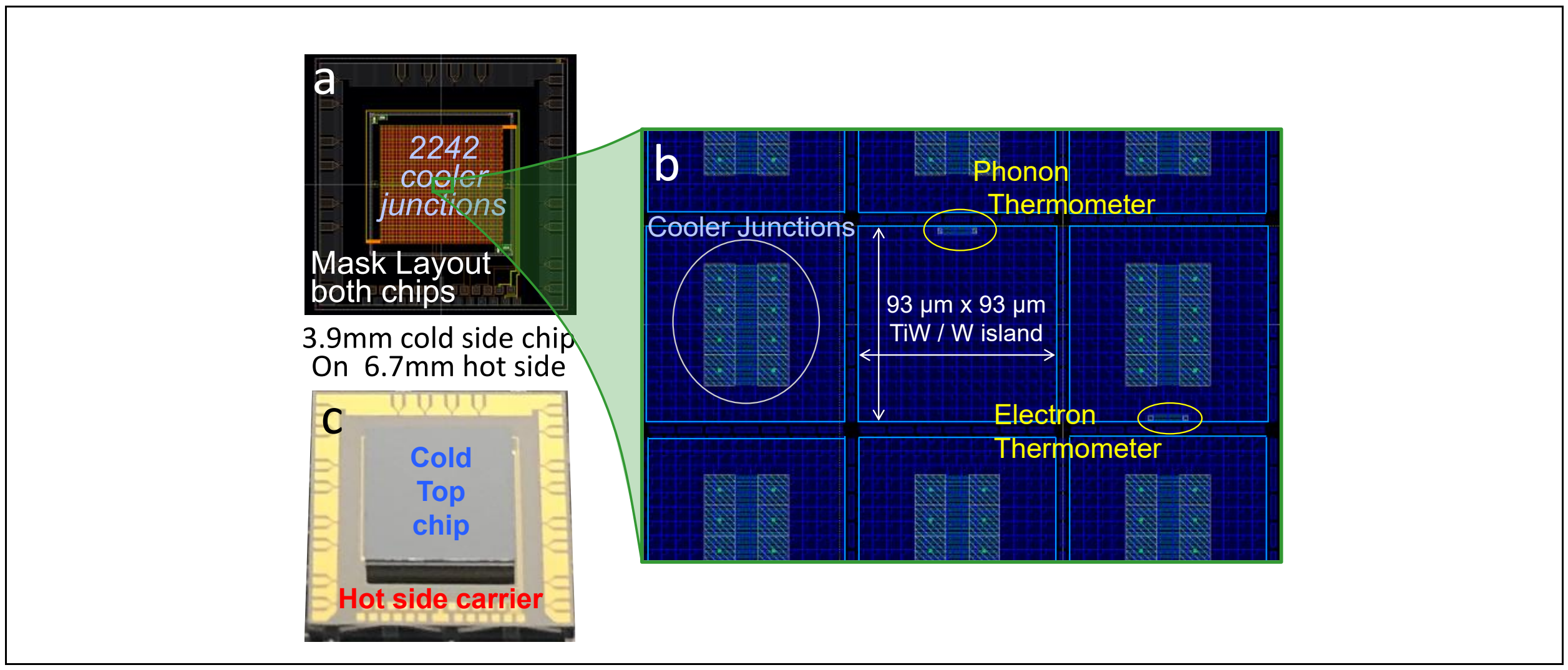

Figure 1 – (a) Lithography mask of the warm side and cold side chips, (b) Zoom-in view of the mask showing normal-metal islands and the SINIS junction pairs and the central normal-metal used as a phonon thermometer next to the cooler with the electron thermometer, (c) Photograph of a flip-chip bump-bonded cooler.

The cooler consisted of 2242 series connected NIS junctions, arranged in pairs (constituting one SINIS junction) such that 1121 normal-metal islands, each measuring 93 µm by 93 µm, are cooled by the same current. Each 786 $\mu m^2$ NIS junction has four 2 µm by 2 µm bumps electrically connected in parallel. One of the SINIS junction pairs also has additional bumps that permit the measurement of the electron temperature in that normal-metal island. The entirety of the cooler array is comprised of 33 rows by 34 columns of SINIS unit cells. In the centrally located row 17 column 17, there is a normal-metal island surrounded by dielectric and therefore not galvanically connected to the cooler current. However, it is bump-bond connected, allowing for a phonon temperature measurement. In this arrangement, the phonon thermometer is actually measuring the electron temperature of this island, but the relatively large normal-metal island volume and electron-phonon coupling result in close agreement between electron and phonon temperatures in the island. Note that by necessity the bumps that permit this measurement are heat leakage paths so that the cold side phonon temperature that is reported is likely to be slightly warmer than the actual phonon temperature in the cold side silicon chip. The figure shows two bumps for each thermometer, but this was later doubled to add a third and fourth thermometer, one at the top and one at the bottom of the two normal-metal islands to increase thermometer connectivity yield, which of course also added more localized thermal leakage to the thermometers. The data reported below is for this two-thermometer-per-island configuration. As such, each cold side chip is connected by 8976 bumps to the warm side chip. This makes the NIS refrigerator mechanically robust and scalable.

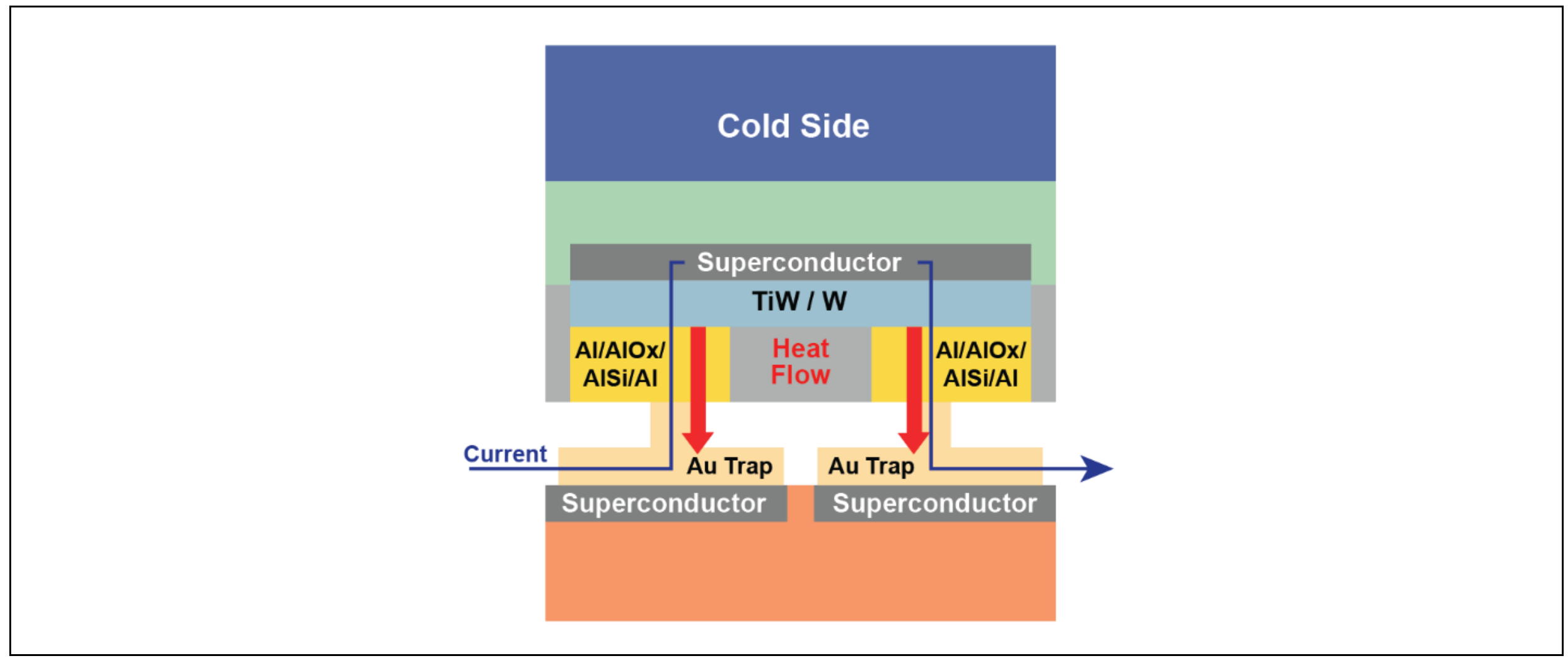

Figure 2 – Cross sectional schematic of a SINIS bump bonded cooler unit cell indicating the current and heat carrying quasi-particle flows.

The cross-sectional schematic of a unit cooler cell which is repeated 1121 times to realize the chip scale refrigerator is shown in Figure 2. Superconducting shunts are used on both cold side and hot side chips to act as current spreader. Current flowing through the normal-metal materials in the cooler is largely restricted to flow primarily through thicknesses and not in-plane, this to reduce Ohmic heating. The presence of the superconducting shunts leads to proximitization of the contacting normal-metal, allowing Cooper pairs to form there in the superconducting mini-gap. On the hot side, a 35-nm thick layer of Cr is used between the superconducting shunt and the Au of the trap to prevent proximitization. For the cold side, the use of Cr was not permissible in the IC foundry, and our earliest junctions showed superconductor-insulator-superconductor (SIS) behavior instead of the desired NIS. This is a clear sign that the TiW / W bilayer was being proximitized with a mini-gap that allowed the current to flow as Cooper pairs on both sides of the $AlO_x$ junction. This was mitigated by using an 80 nm thick Al superconductor shunt with a 400 nm nominally thick TiW layer plus a 200 nm nominally thick W layer. Specific junction resistances were 621 ohm-µm$^2$ for the devices reported in this paper.

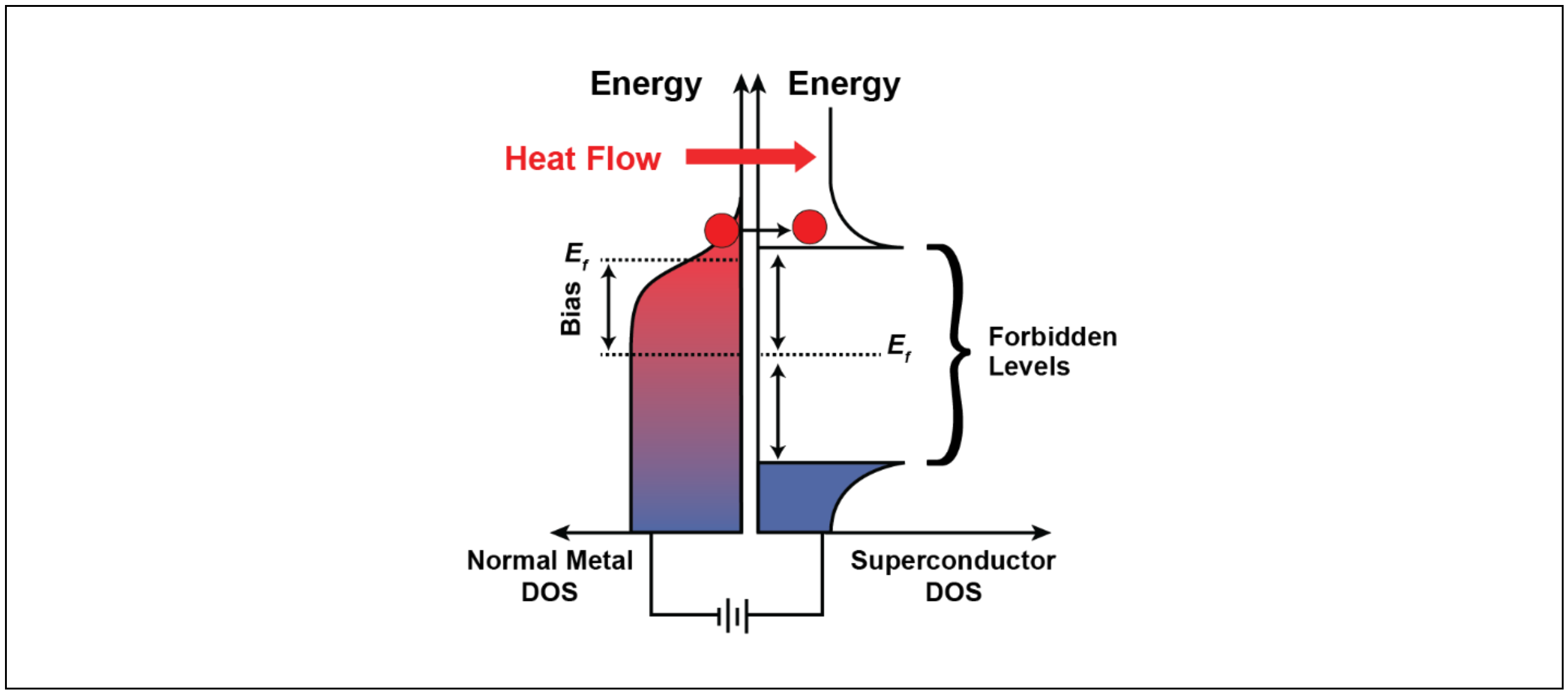

Figure 3 – Illustration of the cooling mechanism in an NIS tunneling junction from hot electrons

Cooling is accomplished by the tunneling of electrons from the normal-metal across the $AlO_x$ junction and into the Al superconducting counter-electrode, taking advantage of the fact that quasiparticle like electrons and quasiparticle like holes carry heat in the same directions from the cold side to the hot side.[14] This is illustrated in Figure 3, where the hot electrons tunnel into the empty states on the superconducting side. Equation 1 gives the approximate expression for the cooling power of an NIS junction assuming only electron tunneling, which doesn't account for Andreev currents, phonons, or any proximity effects.

$$\dot{Q}_{NIS} = \frac{1}{e^2\, R_N}\int_{-\infty}^{\infty} dE\,(E - eV)\boldsymbol{n_S(E)}[f_N(E - eV, T_N) - f_N(E, T_{SC})]$$

$$\boldsymbol{n_S(E)} = \left|\boldsymbol{Re}\left(\frac{(\boldsymbol{E} + \boldsymbol{i\gamma})}{\sqrt{(\boldsymbol{E} + \boldsymbol{i\gamma})^2 - \boldsymbol{\Delta}^2}}\right)\right|$$

Equation 1 –Heat Currents through an NIS junction in the electron tunneling approximation. The superconducting density of states, $\boldsymbol{n_S(E)}$, includes a Dynes parameter $(\boldsymbol{\gamma})$ to account for the lifetime broadening of quasiparticles. The cooling power of an SINIS junction is twice the power of an NIS junction.

Nguyen et al.[15] have shown that direct contact of a normal-metal (in our case this would be the Au of the trap with the Al of the superconducting counter-electrode) is known to inverse-proximitize through to the junction and cause the junction to become Ohmic in nature where the superconducting gap has deteriorated and no longer acts as an energy filter and hence not function as a cooler. The contacting area of the Au of the bump is small but, to ensure that it does not inverse-proximitize the junction, the Al counter-electrode is intentionally fabricated to be 1.4 µm thick. Subsequent use of a Usadel equation solver[16] in a 2D geometry matching the bump and junction areas indicated that this thickness insures a constant gap at the junction surface in our temperature range of interest, 60 to 300 mK.

By limiting the physical connection area between the hot and cold sides the backwards leakage of heat is engineered to enable colder load temperatures and/or higher rejection temperatures. Our best results were found when the bump was deposited as a "hybrid" stack where, after ion beam cleaning of the Al counter-electrode surface and not breaking vacuum, first a 0.66 µm Al layer followed by a 35 nm Ti adhesion layer and then 0.33 µm layer of Au to form a ~1 µm tall bump. This ensures that the hot side to cold side contact is limited in area. Heat leakage occurs at the bump contact through hot side phonons and, for temperature regimes at the contact that are above ~20% of the superconductor's transition temperature Tc, through hot-side electrons from the normal-metal of the trap.

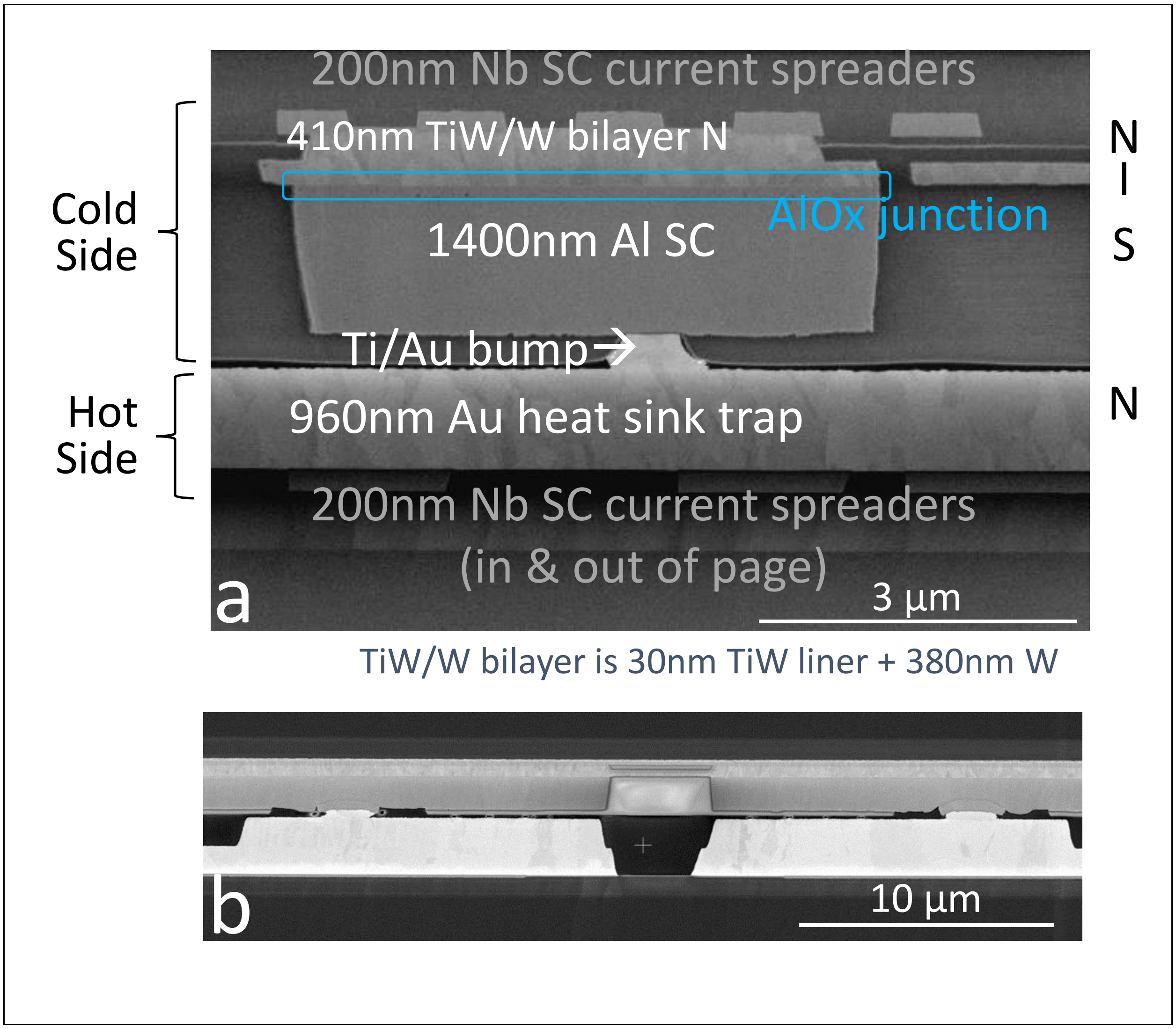

Figure 4 – (a) FIB cross section of a thermometer bump showing the bonded cold side to hot side chip. (b) Argon blade cross section of two neighboring coolers showing the resulting compression of the bumps resulting in 100 nm vacuum gap.

Figure 4a shows a focused ion beam cut through a bonded cold-side chip to hot-side chip pair. This is an early version of a thermometer using a smaller width bump than the 2 µm by 2 µm eventually settled upon for the working devices but since the thermometer and coolers are fabricated simultaneously from the same layers, it shows all of the essential features of the device. The cold side TiW/W bilayer normal-metal was 410 nm thick, later embodiments thickened this to a total of 600 nm under the junction. Our selection of the TiW alloy and W bilayer instead of a normal-metal with a lower resistivity and/or higher electron-phonon coupling such as Au was dictated by foundry limitations. In our process the cooling normal-metal needed to be deposited before junction formation, however, Au deposition is only allowed as the final foundry operation. Initially, the superconducting spreaders on both cold side and hot side chips were made from Nb stripes; this was later changed to solid Al. On the hot side the ~1 μm Au was later thickened to a two-layer stack of Au with a total thickness of ~2 μm, this being found to better facilitate the bump bonding process yield. Recall that the bump is deposited on the junction (cold side) chip, close examination in this image shows that what initially was a 35 nm Ti and 1 μm Au (this cut was taken on a chip before development of the Al/Ti/Au hybrid bump) portion of the bump has compressed in height and spread wider on the Au of the trap (hot side) chip. The vacuum gap between the hot and cold sides is ~100 nm. Near-field radiation across this gap was modeled as evanescent waves[17,18] and found for our temperature regime and materials to be negligible compared to the heat leakage through the bump contacts. Figure 4b shows another bonded cooler which has been Argon blade cut to expose two neighboring junctions and their bumps, where again the vacuum gap is ~100 nm. Progressive cuts through the chips showed that this 100 nm gap was maintained across the entire bonded chip pair, providing confidence that bump bonding is a robust and scalable means to repeatedly and reliably connect the cold and hot sides. Illustrating this, one chip pair was cycled three times from room temperature down to mK and back to room temperature, and then removed from the dilution refrigerator and mounted in a second and different dilution refrigerator and cycled a fourth time, and found to perform consistently in each cool down.

Quasi-particles injected at the NIS junction are diffusively driven by their concentration gradient from the junction to the bump, where they deliver the moved heat (along with any $I \cdot V$ work) to the normal-metal trap. Quasi-particle recombination is detrimental to the efficiency of the refrigerator, with each quasi-particle pair that recombines into a Cooper pair creating a $2\Delta$ phonon in the superconductor, i.e. lattice heating. Although Al as the superconductor is a good choice due to its long quasi-particle lifetime[19,20] and large diffusion coefficient[21,22] only a few percent recombination can be tolerated. This effectively sets the dimensions of the individual cooler junction.

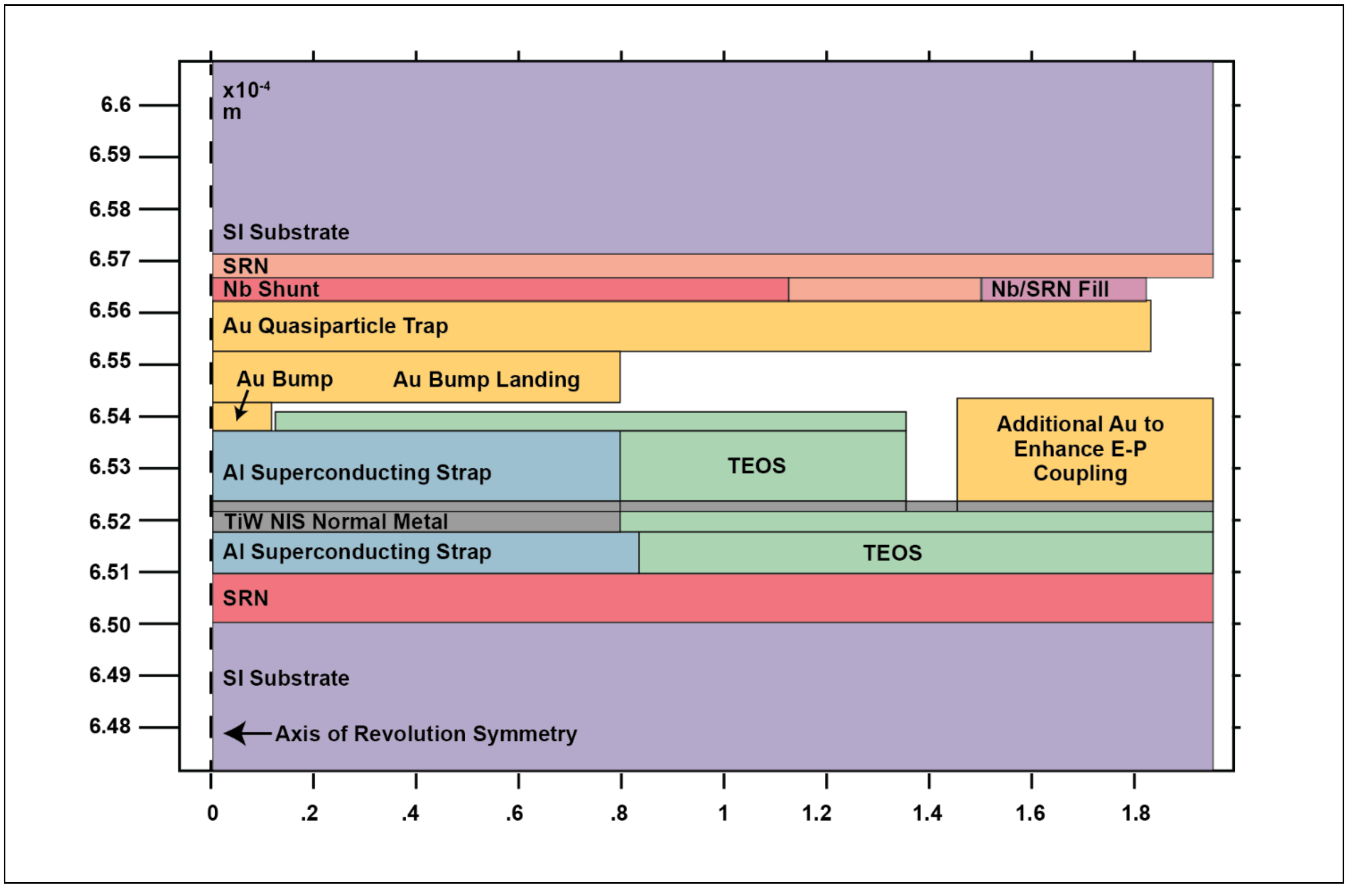

Figure 5 – Cross section of a cooler junction and single bump in a two-dimensional axisymmetric model. (SRN = Silicon-Rich Nitride, TEOS = a-$SiO_2$ derived from TEOS).

## III. MODEL AND RESULTS

COMSOL Multiphysics® version 6.2 2D axisymmetric and 3D models were developed. Figure 5 shows a cross section of a single bump above a junction as a 2D axisymmetric model. In this 2D model interfacial areas and total volumes of each material were set to preserve contact area and total volume associated with ¼ of 1 NIS junction as fabricated. The interface between the Au bump and the Al counter-electrode is of particular importance in controlling the back flow of heat from the hot Au trap to the cold side. The thermal boundary resistance in this model is calculated using the Acoustic Mismatch Model (AMM)[23] taking the Au/Al interface to have a coefficient of 3.925 $K^4cm^2/W$. (This omits consideration of the 35 nm thick Ti adhesion layer.) The COMSOL® model sweeps in current, with the bias voltage back-solved using the standard NIS tunneling current as a function of the integrated energy for each bias voltage[1] which generates a look-up table. The heat lifted by the quasi-particles is calculated using the polylog function analytical heat lift equation of Anghel and Pekola[24]. An example simulation result is shown in Figure 6 where it is seen that, with the exception of the bump interface all of the individual layers are relatively isothermal internally in both lateral and vertical directions, with temperature jumps at layer-to-layer interfaces. The dominant temperature change, as intended by design, is at the narrow bump to counter-electrode interface region. Further improvements towards decreasing the phonon backflow here can be engineered by reducing the bump size, and by introducing superconducting nano-laminates[25] which have been shown to greatly impede phonon flow.

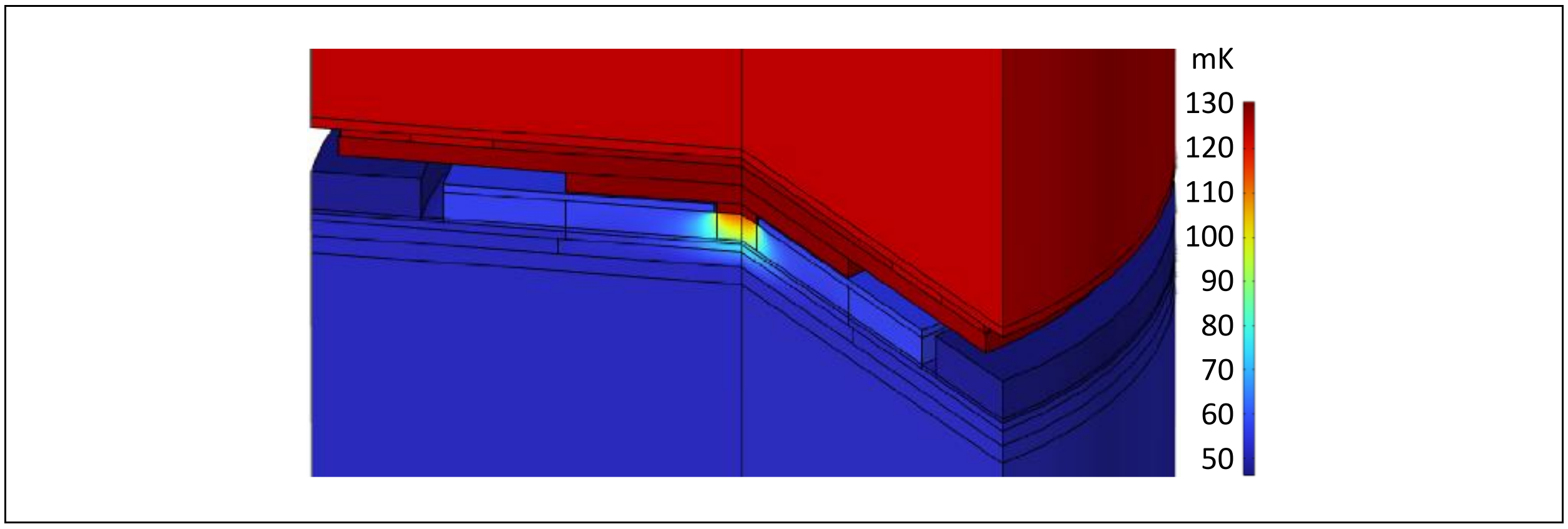

Figure 6 – Phonon temperature profiles at the bump and junction of the 2D model from Figure 5 showing that, as intended, the major temperature jump occurs at the bump. 120 mK rejection, 50 nA (100nA per junction) bias. Note that this model includes additional normal for enhanced electron-phonon coupling, while the comparison to experiment below does not.

Given a typical normal-metal island in our layout contains eight bumps that can be reduced by symmetry to a two bump model, a fully three dimensional COMSOL model was developed. Figure 7 shows the streamlines of the quasiparticles. The injected current is constant over the face of the NIS junction and illustrates how their concentration gradient drives them to the bump contact. Considering the transport distance, quasiparticle branching mixing[26] effects have been ignored (but can be included in future models). Quasiparticle trapping is not treated here, instead it is assumed that any recombination event puts the heat into the superconducting counter-electrode. In the present design our modeling indicates that less than 1% of the injected quasi-particles are lost to recombination into Cooper pairs. The concentration and energy balances follow that as suggested by previous investigators[27,28].

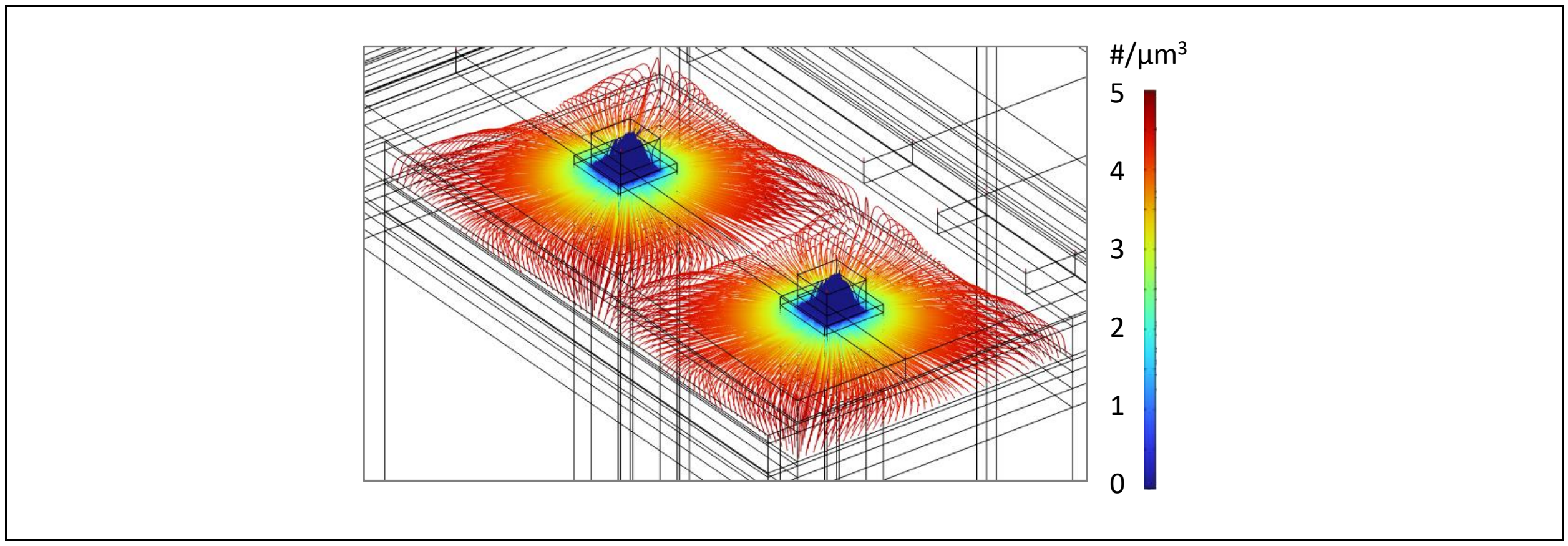

Figure 7 – Quasiparticle density streamlines as calculated using the 3D COMSOL model. 120 mK rejection, 70 mK normal metal temperature, 50 nA per symmetric ½ NIS junction (100 nA per full junction) bias, 2 µm x 2 µm bumps

Andreev currents and heat flows were modelled using the theory of Hekking and Nazarov[29] as elucidated by Rajauria[30] and co-workers[31], and Lowell et al.[32] The voltage biases used in the present experiment generates Andreev currents and heat flows that are orders of magnitude smaller than the tunneling current and heat flow in the coolers. However Andreev currents are present and observable in the thermometers at temperatures below 100 mK at very small voltage biases, which will be presented in a later section.

The two-bump COMSOL model was swept over a range of bias currents initially considering heat leakage by phonon transport only, but found to not match well with the experimental data. The discrepancies appear to be in the amount of backwards heat leakage that occurs at the hot Au trap to Al superconducting counter-electrode interface. Considering only the phonon flow while adjusting the AMM coefficient to different values could not match up with the shape of the experimental data curves. Adapting[33] an equation from Bardeen, Rickayzen, and Tewordt (BRT)[34] relating the ratio of the thermal conductivity of the superconducting state to that in the normal state considering the electrons as heat carriers and modifying that equation to interfacial heat flow resulted in an equation for heat flow across a superconductor to normal-metal interface when $\Delta/k_B T > 4.6$:

$$\left(R_{th,es}^{BRT} A\right)^{-1} = G_{th,es}^{BRT}/A \approx 4\left(k_B^2/e^2\right)\left(\frac{T}{R_c^{BRT} A_c^{BRT}}\right) exp(-y)\{1 + y + \tfrac{1}{2} y^2\}$$

Equation 2– Conductance across a superconductor to normal-metal interface as a function of the electrical contact resistance area product *RcA.*

In Equation 2 $G_{th,es}^{BRT}/A$ is the interfacial thermal conductance [W/m$^2$-K] (the inverse of the interfacial thermal resistance $R_{th,es}^{BRT} A$), $k_B/e$ is Boltzmann's constant divided by the elementary charge, $y = \Delta/k_B T$, $\Delta$ the superconducting gap as a function of the temperature, $T$ the superconducting transition temperature, and $R_c^{BRT}$ [$\Omega$] is the *electrical contact resistance* across a contact area $A_c^{BRT}$ [m$^2$]. When Equation 2 is used in parallel with AMM interfacial phonon conductance, using $(R_c^{BRT} A_c^{BRT}) = 0.157$ [$\Omega\mu m^2$], the COMSOL model's predicted values and curve shape agree better with the data in Figure 8. Equation 2 is useful when the temperature is well below the superconducting transition temperature and the temperature difference is small; larger temperatures and differences will require inclusion of rectification effects.[35]

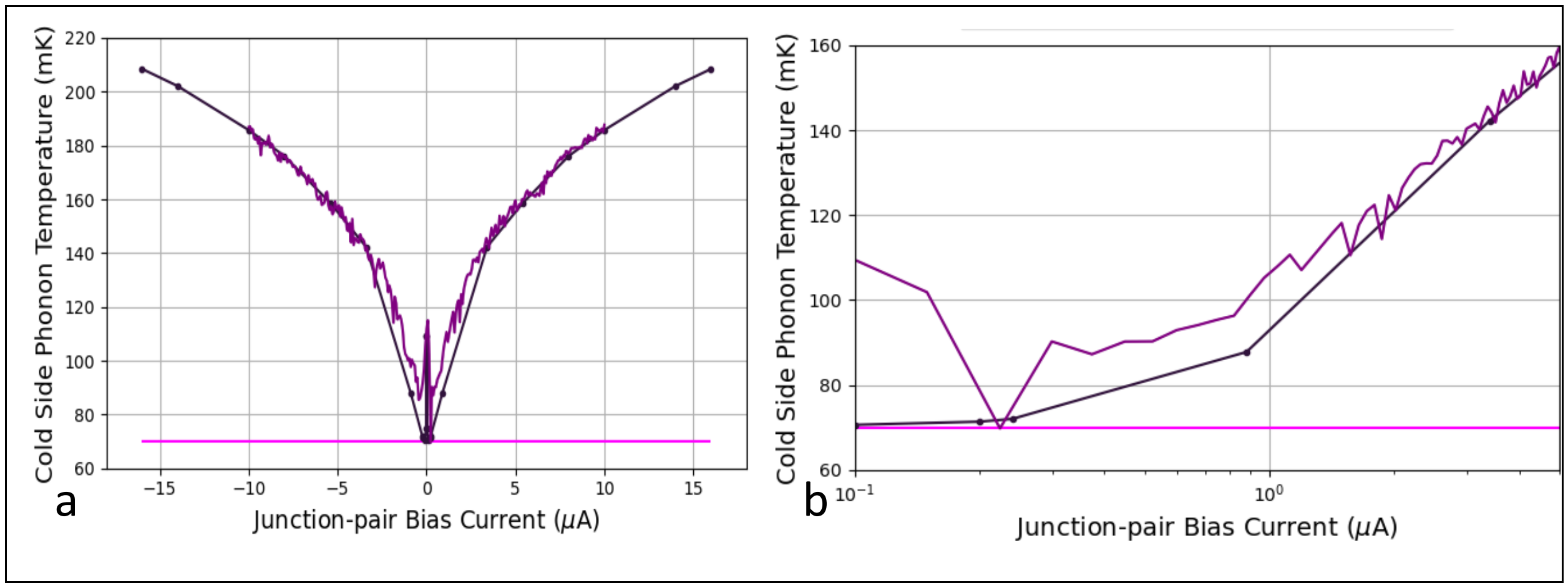

Figure 8 – (a) Phonon cold side temperature as a function of a cooler junction pair (SINIS) bias current at a bath temperature of 120 mK, comparing experimentally measured data (dark purple) with the 3D simulation (black) considering both backwards heat leakage due to phonons (AMM) and electrons (Equation 2) In addition to heating from quasiparticle recombination in the superconductor, (b)

enlargement of the low positive current regime on a log scale for the x-axis The horizontal light purple line is a guide to the eye indicating the coldest phonon temperature measured.

Another important material parameter for predictive modelling of NIS coolers is the electron-phonon coupling coefficient $\Sigma$, whose product with the refrigerated side's normal-metal's volume $V$ determines the amount of heat[36] $Q_{e-p} = \Sigma V\left(T_e^n - T_p^n\right)$ that can be removed from this part of the system. The TiW normal-metal used in our cooler has been measured[37] to have $n \approx 5$ with $\Sigma = 0.5$ nW µm$^{-3}$ K$^5$. This is assumed to also be the values for the tungsten portion of the normal-metal.

While NIS thermometers are capable of being an absolute thermometer[38] our attempts to do so using I-V measurements and discrete derivatives thereof to produce dI/dV, even with filtering failed presumably due to significant electrical noise in the lines that prevent this in our set up. SINIS secondary thermometer calibration was accomplished with the cooler chain unbiased and thermometer current versus voltage curves taken at various temperatures as determined by the dilutions refrigerator's factory installed resistance thermometer[39]. Temperature measurement when the cooler was running was measured by operating the thermometer at a constant current and measuring the resulting voltage across the SINIS thermometer.

Figure 9 shows the electron thermometer response for a series of bath temperatures as the cooler chain bias current is swept (x-axis). On these plots the thermometer current is held as a constant 0.04 µA, such that colder measured temperatures result in higher values on the y-axis. Temperature scales inversely with voltage and the temperature indicated above each plot is the bath temperature. At the base temperature (upper left-hand plot) starting from zero cooler bias current ⓪, the temperature actually increases ① due to the heating caused by the small Andreev (pair transport) current and then turns around when the single particle tunneling starts to remove some heat. However, at this very cold temperature the single particle tunneling is insufficient to cool the normal-metal electrons such that ② is still hotter than the initial ⓪ value. Beyond this, due to the cumulative effect of the raised Fermi level no longer significantly favoring thermally hot electron tunneling, increased I-V power dissipation, and increased quasiparticle recombination events in the superconductor which drives up quasiparticle density, the single particle tunneling results in heating, ③. In the plot of measurements at 100 mK the initial Andreev current is still present ④ but cooling below 100 mK can be achieved. Such a double-inflection curve has been observed at the lowest temperatures in electron cooling using SINIS junctions previously.[39] At 150 mK ⑤ only cooling can be observed, such that no Andreev current can be detected in the electron thermometer in our set up.

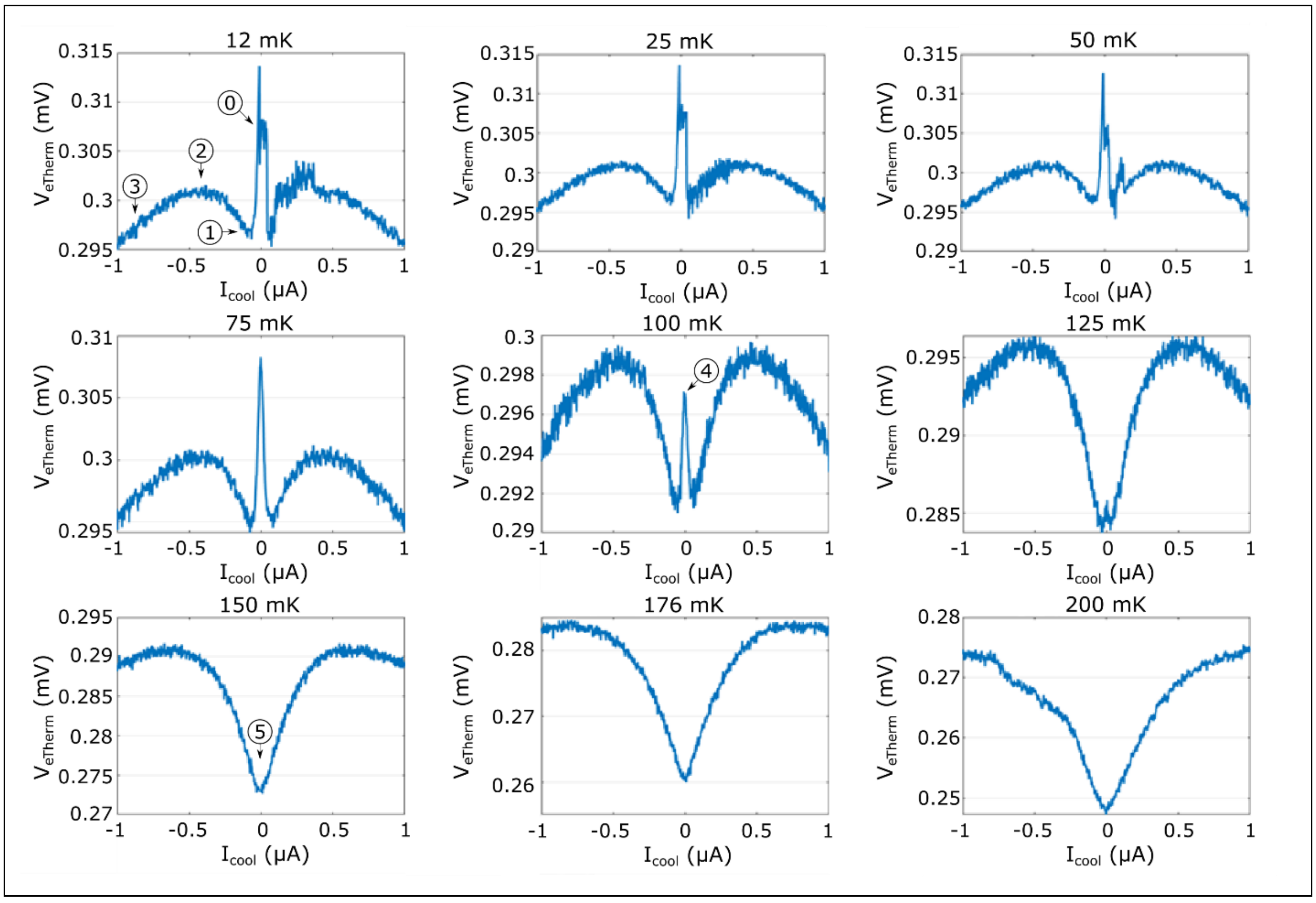

Figure 9 – Electron thermometer measurements at a series of bath temperatures as the cooler chain bias current is swept. Voltages greater than the initial zero bias current point indicate a cooling effect. Temperatures at or below 100 mK show the presence of an Andreev (pair) current. The circled points are described in the text.

Figure 10 is a similar set of sweeps to those of Figure 9 but for the phonon thermometer. Here the measurement is still that of electrons, but in a floating non-galvanically connected piece of normal-metal surrounded by dielectric. Those measured electrons will be well coupled to the phonons. At the coldest mixing chamber temperatures, 50 mK and below, these traces show superconductor – insulator - superconductor (SIS) behavior ⑥ instead of NIS. Even though this thermometer is not galvanically connected to the cooler chain, superconducting strapping was included under the TiW/W normal-metal, and the SIS is attributed to proximitization through the thickness of the normal-metal up to the $AlO_x$ junction insulator of the coolers as a mini-gap which allows Cooper pairs to be on both sides leading to SIS. To investigate this, x-ray diffraction was performed on separate but identically deposited TiW/W bilayer films, which determined it to be alpha-phase body centered cubic, which is the same as bulk tungsten and should have a superconducting transition temperature of 15 mK[40] and as such below our coldest experiments. For contrast we note that beta-phase tungsten films can superconduct up to 130 mK[41]. The expected NIS behavior is observed at higher temperatures. In the 75 mK plot an Andreev current is present ⑦. In the 125 mK plot there is still some small initial Andreev current, but cooling ⑧ is achieved as indicated by the thermometer voltage increasing above the initial zero cooler chain bias current.

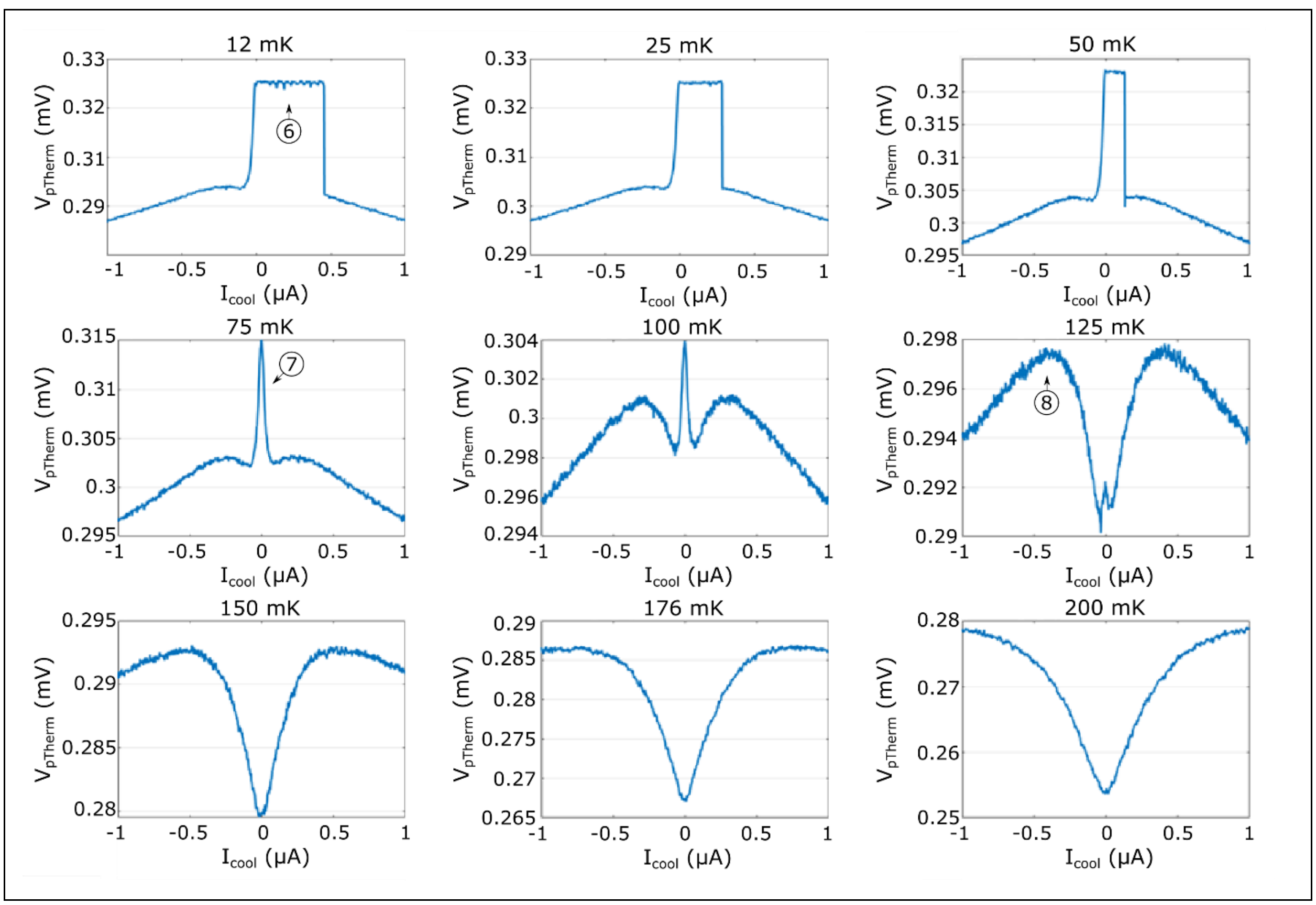

Figure 10 – Phonon thermometer measurements at a series of initial temperatures as the cooler chain bias current is swept. Voltages greater than the initial zero bias current point indicate a cooling effect. The coldest temperatures show SIS behavior. The circled points are described in the text.

Note that the heat rejection on the hot side in the present configuration must flow through the bottom silicon chip, which is limited by the phonon transport from the hot Au of the trap across the metal/dielectric interface, and then out one of two paths. We use Au wire bond wires to connect to 21 electrical ground pads, which represent another dielectric/metal interface, located on the topside periphery of the chip, however these pads only use 7.08 mm$^2$ of the surface; the other heat path flows out through the bottom of the hot side 6.7 mm x 6.7 mm chip which is epoxied to an Au-plated Cu package bottom. Both paths present interfacial thermal boundary resistances limiting the heat flow. To measure the thermal boundary resistance a Ruthenium Oxide (RuOx) resistor was epoxied to the periphery of the (larger) bottom chip and calibrated against the mixing plate temperature. Unfortunately, the epoxy used to attach the RuOx also leaked under the cold side chip and thermally shorted the hot and cold side chips so cooler chain performance was not meaningful in this experiment; what was useful was dissipating heat in one of the thermometers by over-biasing it to form a heat source, and then measuring the bottom chip's temperature. Figure 11 shows the bottom chip's RuOx measured temperature at four different bath temperatures as a function of dissipated power. At the colder temperatures combined with higher powers the temperature difference becomes relatively large, so instead of using small temperature difference $\Delta T$ form of the heat flow $Q = A\,\Delta T/R_b = A\,\Delta T\,T^3/B$ where $A$ is area, $R_b$ the thermal boundary resistance, and $B$ [K$^4$cm$^2$/W] the coefficient of thermal boundary resistance[42] we must use the form $Q = \sigma A(T_{hot}^4 - T_{cold}^4) = \left(\frac{A}{4B}\right)(T_{hot}^4 - T_{cold}^4)$ where $\sigma$ is the Stefan-Boltzmann constant for the phonons[23]. While it is not possible to separate out the contributions from the wire bond pads and the backside epoxy, Figure 11 shows that the combined thermal boundary resistance data at the two coldest temperatures fits well to $\sigma A$=1.80 [mW/K$^4$] and while the temperature rise is considerably smaller at the two higher bath temperature data sets also provides an adequate fit although the data is much noisier.

At high power operation the cooler thus rejects heat to a temperature hotter than the bath temperature.

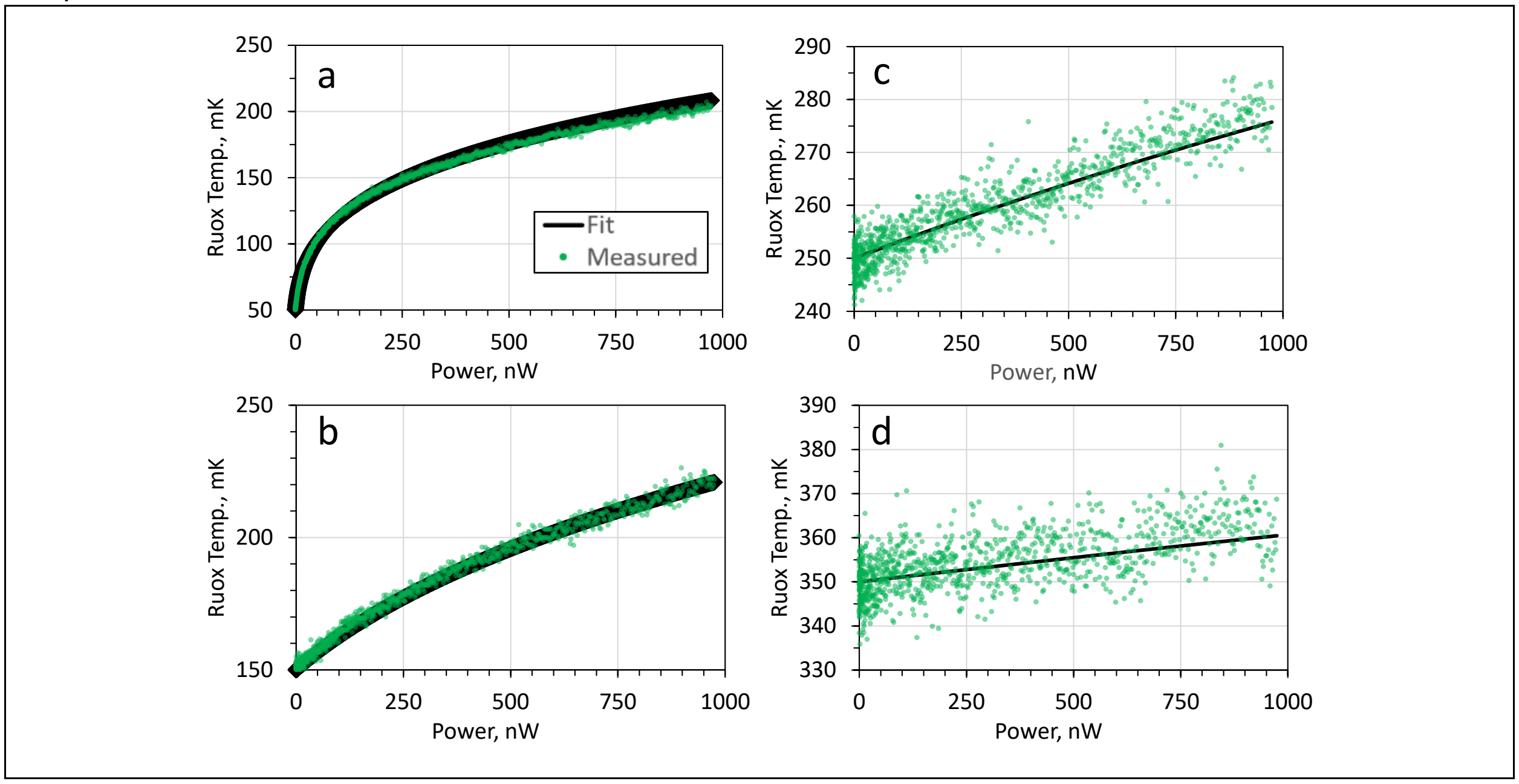

Figure 11 – Hot side chip temperature as measured by a ruthenium oxide thermometer as a function of input power applied to the upper chip at four different bath temperature, showing the excellent fit to a T-fourth law dependency. The bath temperatures are (a) 50 mK, (b) 150 mK, (c) 250 mK, (d) 350 mK.

A better heat path would obviate the trap metal-to-Si dielectric interface and instead allow the hot trap electrons to flow directly to a large normal metal at electrical and thermal ground. In our present design, each of the more than a thousand normal-metal traps is at a different bias along the cooler chain. A large single NIS junction covering many square millimeters or centimeters running the current through a single very large junction would permit this, as it would enable operation with a single bias between the junction cold side and the thermal sink. The power supply in such a case would need to flow several amperes worth of current at a sub-millivolt bias.

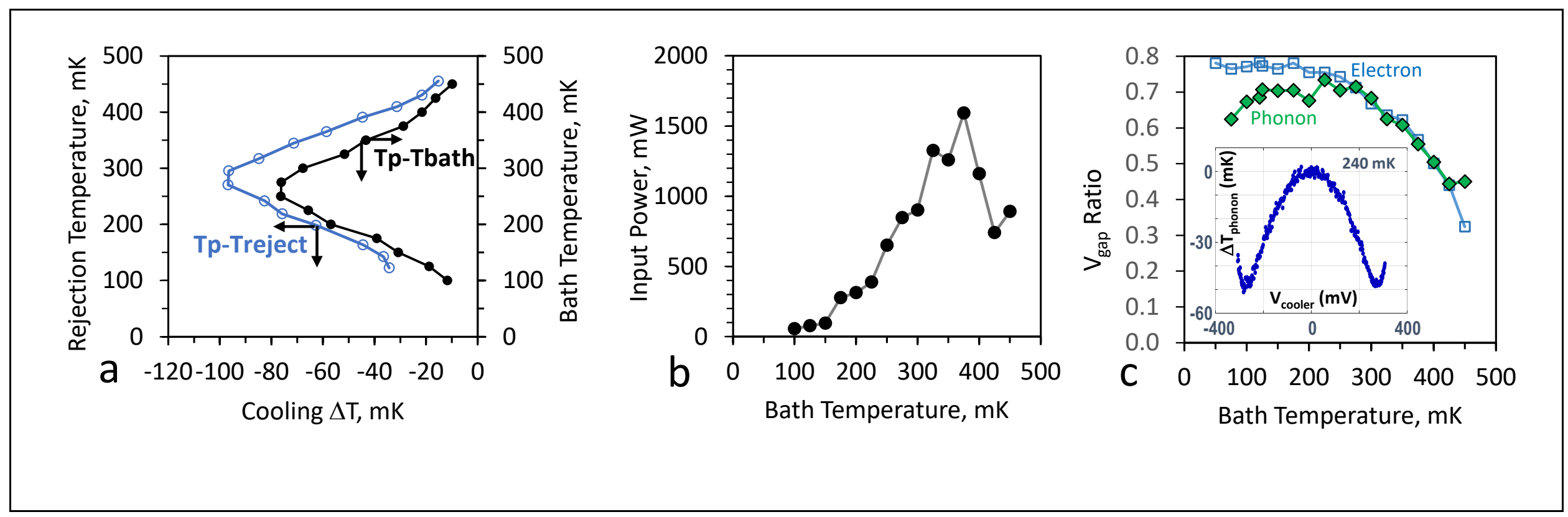

Figure 12 – (a) effective cooling throw, with the cooler chain bias current set to a different optimum at each point based on the phonon thermometer, with the right axis indicating the bath temperature and the left axis the calculated hot side rejection temperature as per the thermal resistances measured in Figure 11, (b) optimal bias power as determined by the coldest temperature achievable at any set bath temperature, (c) optimal cooler bias voltage in terms of $V_{bias}/\Delta_0$. The inset shows an example of the phonon temperature drop as a function of the measured bias voltage on the cooler chain at a temperature of 240 mK.

Running the cooler at its largest temperature difference at a set of bath temperatures as measured by the phonon thermometer is given in Figure 12a. The solid points are relative to the bath temperature, and while we did not have a RuOx thermometer on the hot side chips for this experiment, we can use the thermal boundary resistance measured previously to produce a corrected rejection temperature as indicated by the open points. The largest temperature difference for our tests was $\Delta T =$ -97 mK found at a 271 mK rejection temperature. For comparison, Mykkänen et al.'s[10] stated best cooling using junctions, formed by degenerate-doped silicon to superconducting aluminum, was $\Delta T =$ -83 mK at a bath temperature of 244 mK for their 1 mm diameter silicon island suspended on micro-bridges. Luukanen et al.[6] measured $\Delta T =$ -97 mK at a bath temperature of 220 mK with their SINIS refrigerator on a silicon nitride membrane suspended by micro-bridges. Our values are comparable to the best NIS coolers using micro-bridges or membranes but, by bump-bonding, we have achieved this in a robust and scalable form.

There is no single V-I curve corresponding to the data plotted in Figure 12a, each point at any bath temperature is taken at a different cooler voltage and current, and is optimized separately for the measured electron temperature and measured phonon temperature. The bias power and bias voltage used at each optimized cold side phonon temperature setting is shown in Figure 12b and Figure 12c respectively. The largest input powers occur at the largest bath temperatures, again to be expected since more electrons are available as "hot" (above the Fermi energy) electrons for pulling heat out of the normal-metal.

Heat lift measurements were conducted by over-biasing one of the electron thermometers, turning it into an Ohmic heater and adjusting the cooler bias current until the cold side phonon thermometer measured the same temperature as the bath, this nulls out the thermal gradient between hot and cold sides. Later testing showed that at the high power levels used to drive the cooler, the hot side chip was warmer than the bath (indicated in Figure 11). Thus we can only provide a limited estimate of the cooler's efficiency, but we can use the data obtained to produce a type of a load curve. Figure 13a gives the load (with the over-biased electron thermometer used as the load), the cooler input power at that load (electrical work in), and the calculated hot side chip (rejection) temperature. Recalling that this experiment sought points where the measured phonon temperature equaled the bath it is possible to know the heat lift for various cold side phonon temperatures. The hot side chip's temperature can be calculated using the total power that goes into that chip, namely the load plus electrical work in. Figure 13b shows the calculated Coefficient Of Performance (COP, heater load divided by electrical work in, W/W) and the 2$^{nd}$ Law Efficiency (COP of the tested refrigerator divided by the COP of a Carnot refrigerator). Maximal values occur at 325 mK with a 0.239 COP, and at 275 mK with a 2$^{nd}$ Law Efficiency of 2.24%. While solid state coolers do not achieve the efficiencies of vapor cycle refrigerators, our demonstrated COP is comparable to that of single stage Peltier commercial thermo-electrics operating to cool from near room temperature downwards, which typically have a COP in the range of 0.2 to 0.7[43] and 2$^{nd}$ Law Efficiencies of a few percent. Our maximum measured heat lift is 370 nW at 325 mK (with an electrical work input power of 1.55 µW). Noting that our cold side chip is 0.39 cm by 0.39 cm this corresponds to a load side heat flux of 2.43 µW/cm$^2$. The present layout increased the cold side normal metal volume by making the islands 93 µm by 93 µm occupying 100 µm by 100 µm each on the chip. The junctions were smaller in area, 786 µm$^2$ for each junction (1572 µm$^2$ per normal-metal island). The cooler was created as a 33 row by 34 column array, occupying just 74% of the chip, so in principle increasing the junction fill factor closer to 100% would enable a heat lift of 42 µW/cm$^2$.

Equation 1 shows that the heat lift can directly be improved by using lower resistance tunneling barriers. Recently AlOx tunneling barriers with a specific resistance of 17 ohm-µm$^2$ have been demonstrated in an NIS refrigerator by Hätinen et al.[44] Implementing this value in our cooler would increase the tunneling heat moved by 36x, raising the projected heat lift to above 1 mW/cm$^2$.

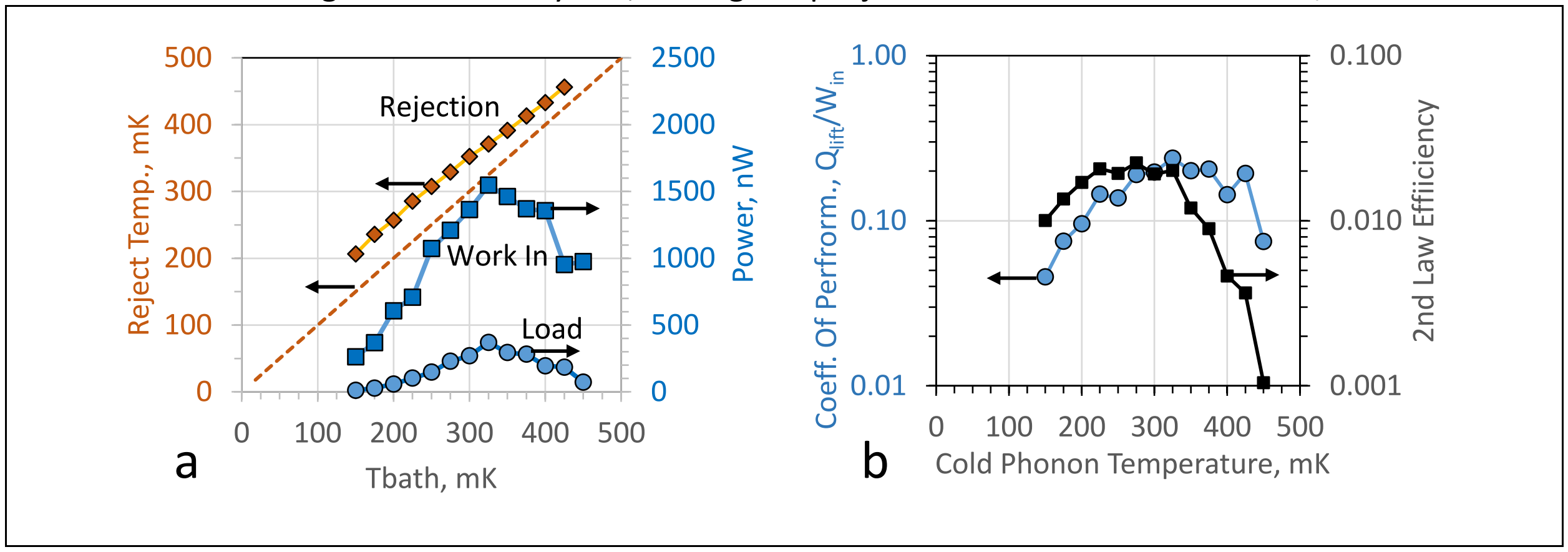

Figure 13 – (a) Efficiency of the cooler as a function of the cold side phonon temperature, (b) Load power, cooler work in, and the calculated hot side (rejection) temperature.

The reported temperatures are for the SINIS thermometers on the cold-side chip, which are connected to the hot-side chip by a pair of Au bumps the same size as used to connect the coolers, so are also likely to measure slightly warmer than elsewhere on the cold-side chip as they act as a small, localized, thermal short. Nonetheless, it is observed that the electrons are cooled more than the phonons, as expected since phonons are cooled through electron-phonon coupling with colder electrons. This is especially true since the volume of electron-phonon coupling normal-metal in each cooler is limited. The maximal throws for the phonons and electrons are also located at slightly different bath temperatures, perhaps due to the significant difference in temperature dependence of electron and phonon thermal conduction paths and heating mechanisms in volumes of relatively poor electron-phonon coupling.

Running a current sweep across the cooler chain and measuring the resulting bias enabled the discrete differentiation $dI/dV$ of the current relative to the voltage which exhibited a BCS peak at 309.0 mV. To determine the superconducting gap of our aluminum we use the thermometers which have small 1 µm$^2$ junctions connected to large normal-metal volumes. Taking the $dI/dV$ we verified the gap to be 0.172 meV, similar to that of bulk aluminum's zero Kelvin 0.175 meV value with a superconducting transition temperature of 1.14 K. Multiplying this by the number of junctions in the cooler chain (2242), can determine the yield of working junctions to be 309.0 mV/(2242*0.172 mV) = 80.1 %. Presumably, the 19.9 % shortfall is due to shorted junctions.

## SUMMARY

We have designed and demonstrated a chip-sized solid-state refrigerator that uses flip-chip bump bonding to allow the flow of heat-carrying quasi-particles driven by a concentration gradient from the cold side to the hot rejection side, limiting the backwards heat leakage by a reduction in the contact area and taking advantage of the temperature jump at a superconductor to normal-metal interface. The maximum measured heat lift was 1.55 µW at 325 mK, the maximum Coefficient Of Performance was 0.239 at 325 mK, and the maximum 2$^{nd}$ Law Efficiency was 2.24% at 275 mK. The maximum temperature throw was -97 mK producing a cold side phonon temperature of 250 mK. Our robust and scalable design is seen to match the best of the membrane and micro-bridge supported SINIS coolers.

A 3D simulation model was developed and matches this cooling data as a function of drive current. Beyond improvements implementing greater fill factor and lower junction resistance, the model of the device reported here indicates that (1) even colder phonon load-side temperatures can be achieved by increasing the normal-metal volume and/or its electron-phonon coupling coefficient (replacing the TiW/W with e.g. Cu or Au), (2) hotter rejection-side temperatures can be attained by increasing the electrical contact resistance to reduce electronic heat transmission from the hot-side normal-metal trap back into the superconductor, and that (3) hotter rejection-side temperatures can be achieved by increasing the thermal resistance at the bump area interface by engineering the acoustic properties of the two sides to reduce phonon transmission. Further optimization of these points should increase the temperature throw to several hundred mK.

Author Contributions

Robert M. Young: initial conception, writing – original draft, thermodynamics, Equation 2
Zachary Stegen: device layout, lithography mask layout, cryogenic testing, formal analysis
John X. Przybysz: superconducting conceptualization, formal analysis
Edward R. Engbrecht: lead IC fabrication
Aaron A. Hathaway: initial COMSOL simulation implementation
Justin C. Hackley: device fabrication, bonding
Kirby B. Myers: COMSOL simulation, incorporating Equation 2
Christian C. Thorpe: COMSOL simulation, incorporating heat flow as a function of bias voltage, near field radiation, and Andreev currents
Aurelius L. Graninger: superconducting materials characterization, NIS thermometer characterization
Robert Miller: RF filtering, Andreev current simulation
Diego A. Morales: test, formal analysis
Roberto D. Carcamo: test, formal analysis
Glen Walters: IC fabrication
Jeric P. Sarad: IC fabrication
Nicholas F. Pleim: Andreev current simulation
Seth Whitsitt: Andreev current equation implementation
Joshua T. Shipman: proximity effect simulation
Anil Erol: Acoustic Mismatch Model development
Melissa G. Loving: x-ray diffraction materials characterization
Evan Donohue: characterization
Corey A. Kegerreis: IC/device layout
Benjamin Dalfort: formalizing of test procedure and data analysis
Moe S. Khalil: NIS thermometer development
Christopher Pinion: IC process flow and fabrication
Randi Jaramillo: Cooler mask layout
John Milinichik: process control monitor layout
Sandro J. Di Giacomo: wafer fabrication
Thomas Zodda: test code, analysis, test stand noise mitigation and filtering

Nilesh Tralshawala: junction sub-gap resistance measurement
Gregory R. Boyd: initial theory and simulations
Jonathan M. Cochran: test, current as a function of bias voltage equation, formal analysis, supervision
Katherine A. Maddock: formal analysis, supervision
Michael P. De Feo: conceptualization, supervision and strategy
Aaron A. Pesetski: initial conceptualization, supervision
Marc E. Sherwin: initial conceptualization, supervision

## Acknowledgement

*Robert M. Young, Zachary Stegen, John X. Przybysz, Edward R. Engbrecht, Aaron A. Hathaway, Justin C. Hackley, Kirby B. Myers, Christian C. Thorpe, Aurelius L. Graninger, Robert Miller, Diego A. Morales, Roberto D. Carcamo, Glen Walters, Jeric P. Sarad, Nicholas F. Pleim, Seth Whitsitt, Joshua T. Shipman, Anil Erol, Melissa G. Loving, Evan Donohue, Corey A. Kegerreis, Benjamin Dalfort, Moe S. Khalil, Christopher Pinion, Randi Jaramillo, John Milinichik, Sandro J. Di Giacomo, Thomas Zodda, Nilesh Tralshawala, Gregory R. Boyd, Jonathan M. Cochran, Katherine A. Maddock, Michael P. De Feo, Aaron A. Pesetski, and Marc E. Sherwin

Relevant US patents are numbers 11333413, 11600760, 11839165, and 10998485.